\documentclass[twocolumn,aps,pra]{revtex4-1}
\usepackage{times}
\usepackage{t1enc}
\usepackage[english]{babel}
\usepackage{graphicx}

\begin{document}

\title{Localization length of nearly periodic layered metamaterials}

\author{O. del Barco and M. Ortu\~no}

\affiliation{Departamento de F\'{i}sica - CIOyN, Universidad de
Murcia, Spain}

\begin{abstract}

We have analyzed numerically the localization length of light
$\xi$ for nearly periodic arrangements of homogeneous stacks
(formed exclusively by right-handed materials) and mixed stacks
(with alternating right and left-handed metamaterials). Layers
with index of refraction $n_1$ and thickness $L_1$ alternate with
layers of index of refraction $n_2$ and thickness $L_2$.
Positional disorder has been considered by shifting randomly the
positions of the layer boundaries with respect to periodic values.
For homogeneous stacks, we have shown that the localization length
is modulated by the corresponding bands and that $\xi$ is enhanced
at the center of each allowed band. In the limit of
long-wavelengths $\lambda$, the parabolic behavior previously
found in purely disordered systems is recovered, whereas for
$\lambda \ll L_1 + L_2$ a saturation is reached. In the case of
nearly periodic mixed stacks with the condition $|n_1 L_1|=|n_2
L_2|$, instead of bands there is a periodic arrangement of
Lorenztian resonances, which again reflects itself in the behavior
of the localization length. For wavelengths of several orders of
magnitude greater than $L_1 + L_2$, the localization length $\xi$
depends linearly on $\lambda$ with a slope inversely proportional
to the modulus of the reflection amplitude between alternating
layers. When the condition $|n_1 L_1|=|n_2 L_2|$ is no longer
satisfied, the transmission spectrum is very irregular and this
considerably affects the localization length.

\end{abstract}

\maketitle

\section{Introduction}

During the last decades, a new type of artificial materials, the
so-called left-handed metamaterials (LH), have attracted a great
deal of attention. They present negative indices of refraction for
some wavelengths \cite{VESELAGO68}, with considerable applications
in modern optics and microelectronics
\cite{YU04,CAL05,MAR08,SHAL07}. Metamaterials can resolve images
beyond the diffraction limit \cite{PEN00,SMITH04}, act as an
electromagnetic cloak \cite{PEN06,LEO06,SCHU06}, enhance the
quantum interference \cite{YANG08} or yield to slow light
propagation \cite{PAPA09}.

Regarding the localization length in disordered systems, the
presence of negative refraction in one-dimensional (1D) disordered
metamaterials strongly suppresses Anderson localization
\cite{MOGI11(a)}. As a consequence, an unusual behavior of the
localization length $\xi$ at long-wavelengths $\lambda$ has been
observed. Asatryan \emph{et al.} reported a sixth power dependence
of $\xi$ with $\lambda$ under refractive-index disorder
\cite{ASA07,ASA10(a)} instead of the well-known quadratic
asymptotic behavior $\xi \sim \lambda^2$
\cite{SHENG86,SHENG90,STER93,SHENG95}. Recently, Mogilevtsev
\emph{et al.} \cite{MOGI10} have also found a suppression of
Anderson localization of light in 1D disordered metamaterials
combining oblique incidence and dispersion
while Torres-Herrera \emph{et al.} \cite{TORR12} have developed a
fourth order perturbation theory to resolve the problem of
non-conventional Anderson localization in bilayered periodic-on-average structures.
The effects of polarization and oblique incidence on light propagation in
disordered metamaterials were also studied in Ref. \cite{ASA10(b)}.

In this article, we calculate numerically the localization length
of light $\xi$ for a one-dimensional arrangement of layers with
index of refraction $n_1$ and thickness $L_1$ alternating with
layers of index of refraction $n_2$ and thickness $L_2$. In order
to introduce disorder in our system, we change the position of the
layer boundaries with respect to the periodic values maintaining
the same values of the refraction indices $n_1$ and $n_2$. This is
the case of positional disorder, in contrast to the compositional
disorder where there exist fluctuations of the index of refraction
\cite{TORR11}.

Two structures will be analyzed in detail: homogeneous stacks (H),
composed entirely by the traditional right-handed materials (RH)
with positive indices of refraction, and mixed stacks (M) with
alternating layers of left- and right- handed materials. For the
sake of simplicity, the optical path in both layers will be the
same, that is, the condition $|n_1 L_1|=|n_2 L_2|$ is satisfied in
most of the work. These periodic-on-average bilayered photonic systems
have already been studied analytically by Izrailev \emph{et al.} \cite{IZRA09,IZRA12}.
These authors have developed a perturbative theory up to second order in the disorder to
derive an analytical expression for the localization length for both H and M stacks.
In our case, we have obtained two equations for the localization
length $\xi$ as a function of the wavelength $\lambda$ from our numerical results.
For H stacks, a quadratic dependence
of $\xi$ for long-wavelengths is found, as previously reported in the
literature. On the other hand, the localization length saturates
for lower values of $\lambda$. An exhaustive study of $\xi$ in the
allowed and forbidden bands (gaps) of weakly disordered systems
will be carried out. We will show that the localization length is
modulated by the corresponding bands and this modulation decreases
as the disorder increases. For low-disordered M stacks and
wavelengths of several orders of magnitude greater than the
grating period $\Lambda = L_1 + L_2$, the localization length
$\xi$ depends linearly on $\lambda$ with a slope inversely
proportional to the modulus of the reflection amplitude between
alternating layers.

The plan of the work is as follows. In Sec.\ II we carry out an
exhaustive description of our one-dimensional disordered system
and the numerical method used in our localization length
calculations. A detailed analysis of $\xi$ in the allowed bands
and gaps of homogeneous stacks is performed in Sec.\ III where a
practical expression for the localization length as a function of
$\lambda$ and the disorder is derived. In Sec.\ IV we calculate
$\xi$ for mixed stacks of alternating LH and RH layers. A linear
dependence of the localization length at long-wavelengths is found
for low-disordered M stacks. Finally, we summarize our results in
Sec.\ V.

\section{System description and numerical model}

\begin{figure}
\includegraphics[width=.40\textwidth]{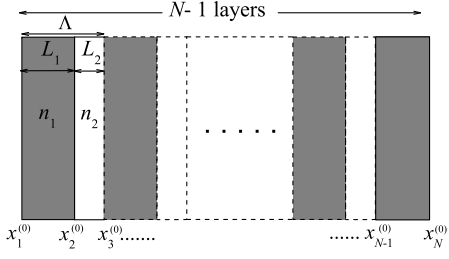}
\caption{A periodic arrangement of layers with index of refraction $n_1$ and
thickness $L_1$ alternating with layers of index of refraction
$n_2$ and thickness $L_2$.
The grating period is $\Lambda = L_1 + L_2$.}\label{fig1}
\end{figure}

Let us consider a one-dimensional arrangement of layers with index
of refraction $n_1$ alternating with layers of index of refraction
$n_2$. The width of each one is the sum of a fixed length $L_i$
for $i=1, 2$ and a random contribution of zero mean and a given
amplitude. The wave-numbers in layers of both types are $k_i =
\omega n_i /c$, where $\omega$ is the frequency and $c$ the vacuum
speed of light. As previously mentioned, the grating period of our
system $\Lambda$ is defined as the sum of the average thicknesses
$L_1$ and $L_2$ of the two types of layers, that is, $\Lambda =
L_1 + L_2$. We have introduced the optical path condition $|n_1
L_1|=|n_2 L_2|$ for simplicity (in the case of left-handed layers
$n_i<0$, so the absolute value has been written to consider these
type of materials). Without disorder, each layer would be limited
by two boundaries $x_j^{(0)}$ and $x_{j+1}^{(0)}$ where $N$ is the
total number of boundaries. The periodic part of the system
considered is schematically represented in Fig.\ \ref{fig1}.

In the presence of disorder, the position of the corresponding
boundaries are
\begin{equation}\label{xjrandom}
x_j = x_j^{(0)} + \xi_j \delta,  \qquad j = 2, \dots, N -1,
\end{equation}
except for the first and the last boundary, so as to maintain the
same total length $L$. The parameters $\xi_j$ are zero-mean
independent random numbers within the interval $[-0.5,0.5]$.
Throughout all our calculations, we have chosen values of the
disorder parameter $\delta$ less than $L_1$ and $L_2$.

For each $L$, we calculate the transmission coefficient of our
structure $T$ and average its logarithm, $\ln T$, over 800
disorder configurations. Then, we obtain numerically the
localization length $\xi$ via a linear regression of $\ln T$
\cite{TORR11}
\begin{equation}\label{linearregr}
\lim_{L\to\infty}-\frac{\left\langle \ln T \right\rangle}{2L} =
\frac{1}{\xi}.
\end{equation}
Here, the angular brackets $\langle ... \rangle$ stand for
averaging over the disorder. We choose 6 values of the total
length $L$ to perform the linear regression of Eq.\
(\ref{linearregr}). The localization length $\xi$ is evaluated as
a function of the disorder parameter $\delta$ and the frequency of
the incident photon $\omega$.

We calculate the transmission coefficient of our system via the
characteristic determinant method, firstly introduced by Aronov
\emph{et al.} \cite{AG91}. This is an exact and non perturbative
method that provides the information contained in the Green
function of the whole system. In our case, the characteristic
determinant $D_j$ can be written as \cite{AG91}
\begin{equation}\label{recrelation}
D_j = A_j D_{j-1} - B_j D_{j-2},
\end{equation}
where the index $j$ runs from 1 to $N$ and the coefficients $A_j$ and $B_j$ can be written as
\begin{equation}\label{Aj}
A_j = 1 + \lambda_{j-1,j} \ \frac{r_{j-1,j}}{r_{j-2,j-1}},
\end{equation}
and
\begin{equation}\label{Bj}
B_j = \lambda_{j-1,j} \ \frac{r_{j-1,j}}{r_{j-2,j-1}} \left( 1 - r_{j-2,j-1}^2 \right).
\end{equation}
The parameters $r_{j-1,j}$, which are the reflection amplitudes
between media $j-1$ and $j$, are given by
\begin{equation}\label{reflecampl}
r_{j-1,j} = -r_{j,j-1} = \frac{Z_{j-1} - Z_j}{Z_{j-1} + Z_j},
\end{equation}
where $Z_{j}$ corresponds to the impedance of layer $j$ and can be
be expressed for normal incidence in terms of its dielectric
permittivity $\epsilon_j$ and magnetic permeability $\mu_j$
\begin{equation}\label{impedancej}
Z_j = \sqrt{\frac{\mu_j}{\epsilon_j}}.
\end{equation}

The quantity $\lambda_{j-1,j}$ entering Eqs.\ (\ref{Aj}) and
(\ref{Bj}) is a phase term \cite{AG91}
\begin{equation}\label{lambdaj}
\lambda_{j-1,j} = \lambda_{j,j-1} = \exp\left[2 i k_{j-1} |x_j -
x_{j+1}| \right].
\end{equation}
Here $k_{j-1}$ is the wave-number in a layer with boundaries $x_j$
and $x_{j+1}$. This recurrence relation facilitates the numerical
computation of the determinant. The initial conditions are the
following
\begin{equation}\label{inconditions}
A_1 = 1; \qquad D_0 = 1; \qquad D_{-1} = 0.
\end{equation}
The transmission coefficient of our structure $T$ is given in
terms of the determinant $D_N$ by
\begin{equation}\label{coefftrans}
T = |D_N|^{-2}.
\end{equation}

\section{Localization length for homogeneous stacks}

Before dealing with mixed stacks, we present results for
low-disordered homogeneous systems with underlying periodicity,
which has not been previously studied. In this section we perform
a detailed analysis of the localization length $\xi$ in the
allowed bands and in the forbidden gaps of disordered H stacks as
a function of the disorder $\delta$, the incident wavelength
$\lambda$ and the reflection coefficient between alternating
layers $|r_{j-1,j}|^2$.

As it is well known, in the absence of disorder the transmission
spectrum of right-handed systems presents allowed and forbidden
bands whose position can be easily determined via the following
dispersion relation obtained from the Bloch-Floquet theorem
\cite{YAR84}
\begin{eqnarray}\label{beta}
\cos (\beta \Lambda) &=& \cos (k_1 L_1)\cos (k_2 L_2)
            \nonumber\\
&-& \frac{1}{2} \left( \frac{Z_2}{Z_1} + \frac{Z_1}{Z_2} \right)
\sin(k_1 L_1) \sin(k_2 L_2),
\end{eqnarray}
where $\beta$ is the Block wave-vector. When the modulus of the
right-hand side of Eq.\ (\ref{beta}) is greater than 1, $\beta$
has to be taken as imaginary. This situation corresponds to a
forbidden band. Taking into account the condition $|n_1 L_1|=|n_2
L_2|$, Eq.\ (\ref{beta}) reduces to
\begin{equation}\label{betasimpl}
\cos (\beta \Lambda) = \cos^2 (k_1 L_1) -
\frac{1}{2} \left( \frac{Z_2}{Z_1} + \frac{Z_1}{Z_2} \right)
\sin^2(k_1 L_1).
\end{equation}
On the other hand, when $\cos (\beta \Lambda)$ is equal to unity,
the incident frequency $\omega$ is located at the center of the
$m$-th allowed band, $\omega_{\rm c}^{(m)}$. After some algebra,
we obtain from Eq.\ (\ref{betasimpl})
\begin{equation}\label{omegacenter}
\omega_{\rm c}^{(m)} = m \pi \left(\frac{c}{n_1 L_1} \right)= m
\pi \left(\frac{c}{n_2 L_2}\right).
\end{equation}

\begin{figure}
\includegraphics[width=.40\textwidth]{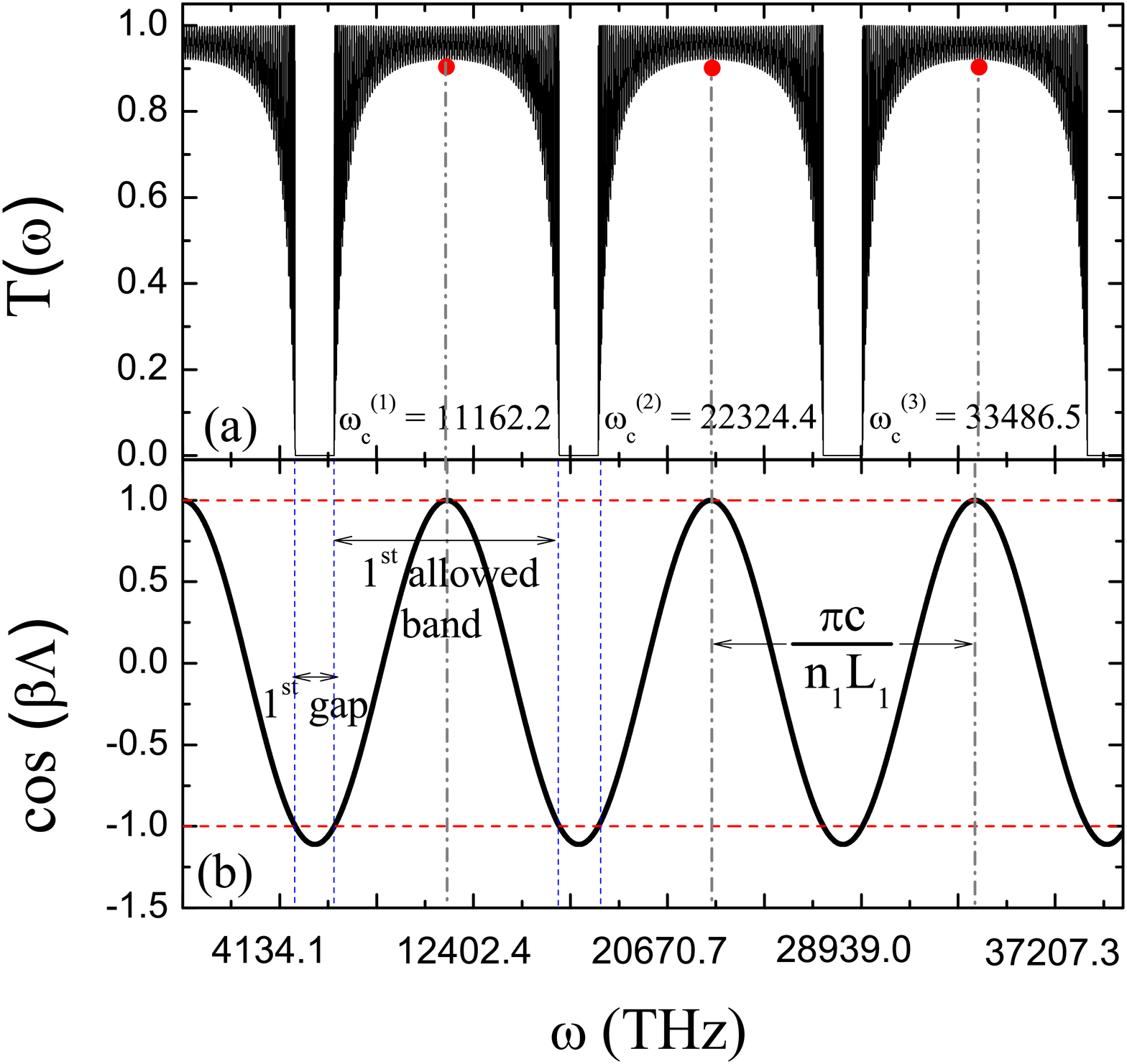}
\caption{(a) The transmission coefficient $T$ and
(b) the parameter $\cos (\beta \Lambda)$ versus the frequency
$\omega$ for the homogeneous periodic system described in the text
(99 layers).}\label{fig2}
\end{figure}

Let us first consider a periodic H stack formed by 50 layers of
length $L_1=$ 52.92 nm and index of refraction $n_1=$ 1.58
alternating with 49 layers of length $L_2=$ 39.38 nm and $n_2=$
2.12. The total size of our structure is 4.57 $\mu$m and the
reflection coefficient between alternating layers 0.05259. Fig.\
\ref{fig2}(a) represents the transmission coefficient $T$ as a
function of the frequency $\omega$ to illustrate its behavior.
Also shown are the center of each allowed band calculated via Eq.\
(\ref{omegacenter}). There are 99 peaks in each band so they can
hardly been resolved on the scale used. Moreover, in Fig.\
\ref{fig2}(b) the parameter $\cos (\beta \Lambda)$ is plotted
versus the frequency $\omega$ for this periodic system. The first
gap and the first allowed band have been shown for a better
comprehension.

\begin{figure}
\includegraphics[width=.45\textwidth]{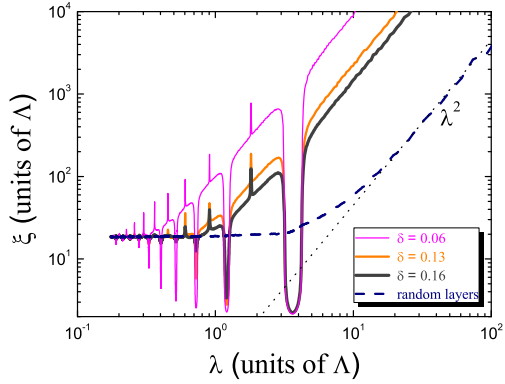}
\caption{Localization length $\xi$ versus the
wavelength $\lambda$ for different values of the disorder
parameter $\delta$. The H stack corresponds to the arrangement
represented in Fig.\ \ref{fig2} but now 50000 layers have been
considered. The dashed line stands for the ''total disorder''
case. All lengths are expressed in units of the grating period
$\Lambda$.}\label{fig3}
\end{figure}

A systematic numerical simulation of a realistic system with 50000
layers has been carried out. The parameters are the same as in the
previous example. In Fig.\ \ref{fig3} we represent the
localization length $\xi$ versus the wavelength $\lambda$ for
different values of the disorder parameter $\delta$ (shown in the
legend of the figure). The dashed line corresponds to ''total
disorder'', that is, an arrangement of layers with random
boundaries and alternating indices of refraction $n_1$ and $n_2$.
Several features are evident in the figure. For long-wavelengths,
one observes a quadratic asymptotic behavior, as can be compared
with the dotted line \cite{SHENG86,SHENG90,STER93,SHENG95}. An
in-deep numerical analysis of the coefficient characterizing this
dependence has been performed. To this aim, 20 different H stacks
were considered and the following expression for the localization
length was found
\begin{equation}\label{longlocHinf}
\xi \simeq 0.063 \frac{\lambda^2}{\Lambda_{\rm op}^2 r^2
\delta^2}, \qquad \textrm{for} \quad \lambda \to \infty
\end{equation}
where $\Lambda_{\rm op} = n_1 L_1 + n_2 L_2$ is the optical path
across one grating period $\Lambda$. All the lengths in Eq.\
(\ref{longlocHinf}) are expressed in units of $\Lambda$. In the
opposite limit of short $\lambda$, the localization length $\xi$
saturates to a constant value \cite{ASA10(a),BALU85}. Our numerical
results have shown that this constant is proportional to the
inverse of the reflection coefficient between alternating layers
$|r|^2$, that is,
\begin{equation}\label{longlocHshort}
\xi \simeq \frac{1}{r^2}, \qquad \textrm{for} \quad \lambda \to 0.
\end{equation}

Izrailev \emph{et al.} \cite{IZRA09,IZRA12} have developed a perturbative theory up to
second order in the disorder to calculate analytically the localization length in both
homogeneous and mixed stacks. This model is quite general and is valid for both quarter
stack medium (mainly considered in our work) and systems with different optical widths.
Assuming uncorrelated disorder and random perturbations with the same amplitude in both layers
(the main considerations in our numerical calculations) one can easily derive the following
analytical expression for $\xi$ at long-wavelengths from Izrailev's formulation
\begin{equation}\label{izragen}
\xi = \frac{Z_1 Z_2}{(Z_1 - Z_2)^2} \ \frac{2\lambda^2}{\pi^2 (n_1^2 + n_2^2) \delta^2}.
\end{equation}
For similar values of the layer impedances $Z_1 \simeq Z_2$, the first term in Eq.\
(\ref{izragen}) can be approximated by $1/4 r^2$ and $n_1^2 + n_2^2 \simeq 2 \Lambda_{\rm op}^2$, so Eq.\
(\ref{izragen}) reduces to
\begin{equation}\label{izradef}
\xi = \left( \frac{1}{4\pi^2}\right) \frac{\lambda^2}{\Lambda_{\rm op}^2 r^2 \delta^2}
\simeq 0.025 \frac{\lambda^2}{\Lambda_{\rm op}^2 r^2 \delta^2}, \qquad \textrm{for}
\quad \lambda \to \infty
\end{equation}
which is similar to our numerical expression Eq.\ (\ref{longlocHinf}).

The randomness only affects partially the periodicity of the
system, which manifests in the existence of bands and gaps. The
localization length depends on the position in the band and on the
disorder. The modulation of $\xi$ by the bands can be clearly
appreciated in Fig.\ \ref{fig3}. These results are consistent with
other published works on this topic \cite{MOGI11(b),LUNA09}.
Recently, Mogilevtsev \emph{et al.} \cite{MOGI11(b)} have reported
that the photonic gaps of the corresponding periodic structure are
not completely destroyed by the presence of disorder while
Luna-Acosta \emph{et al.} \cite{LUNA09} have shown that the
resonance bands survive even for relatively strong disorder and
large number of cells.

\begin{figure}
\includegraphics[width=.45\textwidth]{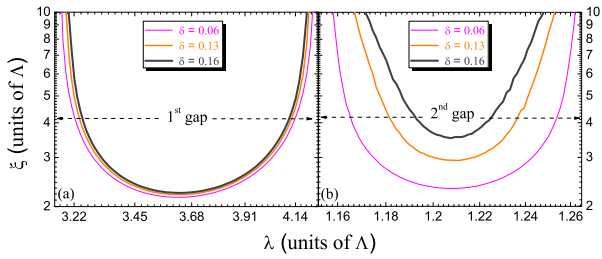}
\caption{Localization length $\xi$ versus the
wavelength $\lambda$ for (a) the first and (b) the second gaps
depicted in Fig.\ \ref{fig3}.}\label{fig4}
\end{figure}

Having a close look into the first gap in Fig.\ \ref{fig3}, one
observes that the localization length is practically independent
of the disorder $\delta$. In order to visualize this effect, Fig.\
\ref{fig4} represents (a) the first and (b) the second gaps
depicted in Fig.\ \ref{fig3}. As mentioned, the dependence of
$\xi$ with the disorder is almost negligible in the first gap.
When the wavelength is similar to the grating period $\Lambda$,
the influence of the disorder is greater, as can be easily deduced
from simple inspection of Fig.\ \ref{fig4}(b).

\begin{figure}
\includegraphics[width=0.47\textwidth]{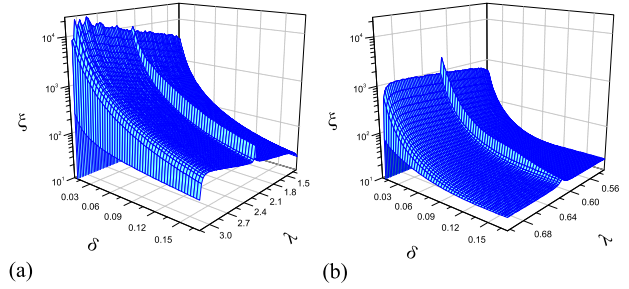}
\caption{Three-dimensional graphs of the
localization length $\xi$ versus the wavelength $\lambda$ and the
disorder $\delta$ for (a) the first and (b) the third allowed
bands. The H stack is the same as in Fig.\ \ref{fig2}. All lengths
are expressed in units of the grating period
$\Lambda$.}\label{fig5}
\end{figure}

Let us now focus on the allowed bands and study in detail the
behavior of the localization length in these regions. To this aim,
three-dimensional (3D) graphs of $\xi$ versus the wavelength
$\lambda$ and the disorder $\delta$ have been plotted in Fig.\
\ref{fig5} for (a) the first and (b) the third allowed bands (see
again Fig.\ \ref{fig2}). All this magnitudes have been normalized
to the grating period $\Lambda$. The localization length $\xi$ is
enhanced in a small region around the center of each allowed band.
A similar result was found by Hern\'{a}ndez-Herrej\'{o}n \emph{et
al.} \cite{HERN10} who obtained a resonant effect of $\xi$ close
to the band center in the Kronig-Penney model with weak
compositional and positional disorder. This increase in the localization
length is due to emergence of the Fabry--Perot resonances associated with
multiple reflections inside the layers from the interfaces \cite{IZRA09,IZRA12,IZRA10}.
In particular, for homogeneous quarter stack systems, the Fabry--Perot resonances
arise exactly in the middle of each allowed band where $\beta$ vanishes \cite{IZRA09,IZRA12}.
The saturation of $\xi$ for short-wavelengths is also appreciated in these 3D images.

\begin{figure}
\includegraphics[width=.49\textwidth]{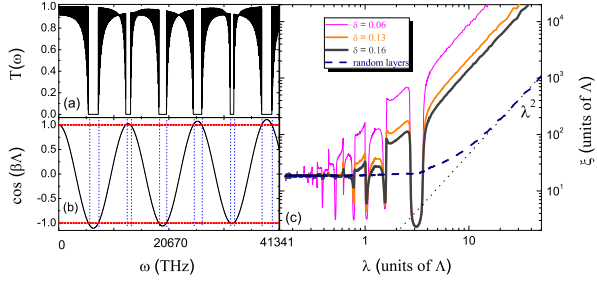}
\caption{(a) The transmission coefficient $T$ and
(b) the parameter $\cos (\beta \Lambda)$ versus the frequency
$\omega$ for the asymmetric periodic H stack described in the main
text (99 layers) and (c) the corresponding localization length
$\xi$ versus the wavelength $\lambda$ for different disorder
parameters $\delta$ (50000 layers).}\label{fig6}
\end{figure}

Up to now, H stacks with the same optical path in layers of both
types have been considered, that is, arrangements verifying the
condition $|n_1 L_1|=|n_2 L_2|$ in the absence of disorder. As a
consequence, the transmission spectrum $T$ of the corresponding
periodic system presented a symmetric distribution of allowed
bands and gaps (as previously shown in Fig.\ \ref{fig2}). What
happens in the case of a non-symmetric band distribution, that is,
when the condition $|n_1 L_1|=|n_2 L_2|$ is not satisfied? To
answer this question, we have plotted the transmission coefficient
$T$ (Fig.\ \ref{fig6}(a)) and the parameter $\cos(\beta \Lambda)$
(Fig.\ \ref{fig6}(b)) versus the frequency $\omega$ for a periodic
H stack formed by 50 layers of length $L_1=$ 52.92 nm and index of
refraction $n_1=$ 1.58 alternating with 49 layers of length $L_2=$
28.80 nm and $n_2=$ 2.12. Note that the condition $|n_1 L_1|=|n_2
L_2|$ is no longer held, so the band structure is asymmetric.
Accordingly, the localization length $\xi$ shown in Fig.\
\ref{fig6}(c) presents an irregular form in the allowed and
forbidden bands. As in the symmetric case, no band modulation
exists for high disorders and the quadratic asymptotic behavior
for long-wavelengths is also verified. Moreover, the peaks in the localization
length due to Fabry--Perot resonances still can be appreciated,
although they are no longer in the center of the bands \cite{IZRA09,IZRA12}.
A total number of 50000 layers was considered in our localization length calculations.

\section{Localization length for mixed stacks}

Once analyzed in detail the behavior of the localization length
$\xi$ for homogeneous systems, let us now deal with M stacks
composed of alternating LH and RH layers.

\begin{figure}
\includegraphics[width=.40\textwidth]{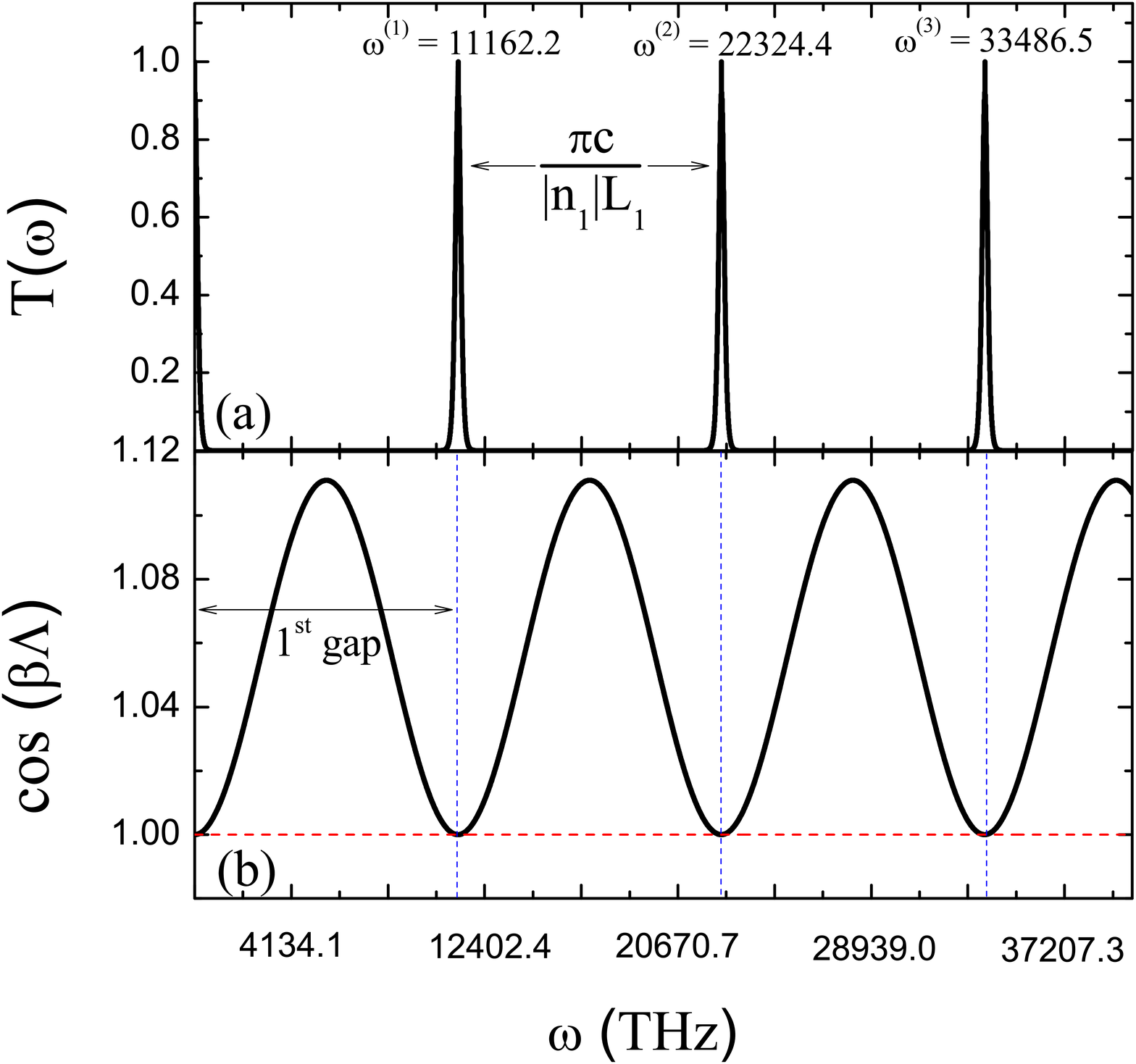}
\caption{(a) The transmission coefficient $T$ and
(b) the parameter $\cos (\beta \Lambda)$ versus the frequency
$\omega$ for the mixed periodic system described in the text (99
layers).}\label{fig7}
\end{figure}

In our numerical calculations we have considered a periodic M
stack formed by 50 layers of length $L_1=$ 52.92 nm and index of
refraction $n_1=$ -1.58 alternating with 49 layers of length
$L_2=$ 39.38 nm and $n_2=$ 2.12. Again, the condition $|n_1
L_1|=|n_2 L_2|$ has been imposed. Note that this arrangement has
similar parameters than the one depicted in Sec.\ III, but now
$n_1$ is negative.  This change of sign results in a severe
modification of the transmission coefficient $T$, as we will show
immediately. For this the periodic system, Fig.\ \ref{fig7}
represents (a) the transmission coefficient $T$ and (b) the
parameter $\cos (\beta \Lambda)$ versus the frequency $\omega$ of
the incident light. Unlike the H stack case, no allowed bands
exist and practically the entire transmission spectrum is formed
by gaps. A set of periodically distributed Lorentzian resonances
is found instead. The position of the center of each resonance is
given by Eq.\ (\ref{omegacenter}), that is, the center of the
allowed bands in homogeneous systems.

\begin{figure}
\includegraphics[width=.45\textwidth]{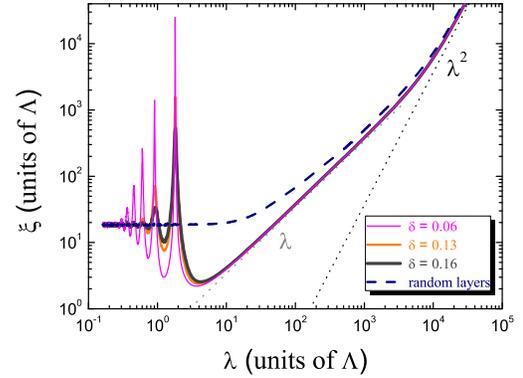}
\caption{Localization length $\xi$ versus the
wavelength $\lambda$ for different disorder parameters $\delta$.
The M stack corresponds to the one represented in Fig.\ \ref{fig7}
but here 50000 layers have been considered. The dashed line stands
for the ''total disorder'' case.}\label{fig8}
\end{figure}

In respect to the localization length, positional disorder was
introduced as explained in Sec.\ II. As previously considered, the
total number of layers in our numerical calculations was 50000 and
the number of disordered configurations to average the logarithm
of the transmission coefficient was 800. The result is shown in
Fig.\ \ref{fig8} where the localization length $\xi$ is
represented versus the wavelength $\lambda$ for different values
of the disorder parameter $\delta$. The dashed line corresponds to
the ''total disorder'' case. Again, for long-wavelengths a
quadratic asymptotic behavior of $\xi$ is found, but now a region
where the localization length is proportional to $\lambda$ exists.
We will turn to this point in the next figure to quantify the
slope of this linear dependence. As it is noticed, the Lorentzian
resonances associated with multiple reflections in the layers
modulate the shape of $\xi$ and this modulation
decreases as the disorder increases. Moreover, the saturation of
the localization length for low-wavelengths can also be
appreciated. As in the H stack case, the constant where $\xi$
saturates is proportional to the inverse of the reflection
coefficient between alternating layers $|r|^2$.

\begin{figure}
\includegraphics[width=.45\textwidth]{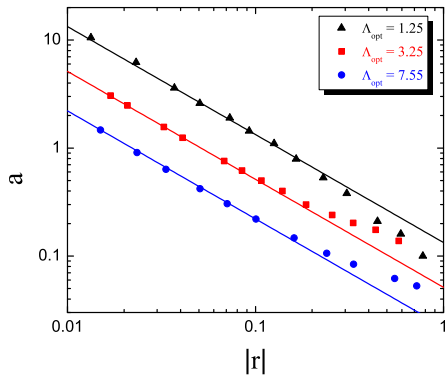}
\caption{Numerical calculations of the slope $a$
versus $|r|$ for several values of $\Lambda_{\rm op}$ (expressed
in units of the grating period $\Lambda$). The solid lines
correspond to the results obtained via Eq.\
(\ref{longlocM}).}\label{fig9}
\end{figure}

The linear dependence of $\xi$ with the wavelength $\lambda$ has
been exhaustively studied by our group to find a simple analytical
expression for the localization length in this region. More than
30 different M stacks have been simulated and we have arrived at
the following empirical equation
\begin{equation}\label{longlocM}
\xi = \frac{\lambda}{6 \Lambda_{\rm op} |r|} = a \lambda,
\end{equation}
where $\xi$, $\lambda$ and $\Lambda_{\rm op}$ are expressed in
units of the grating period $\Lambda$. In Fig.\ \ref{fig9}, our
numerical calculations of the slope $a$ versus $|r|$ have been
plotted for several values of $\Lambda_{\rm op}$, triangles
(1.25), squares (3.25) and circles (7.55). The solid lines
correspond to the results obtained via Eq.\ (\ref{longlocM}). One
notices a good degree of validity for a wide range of $|r|$
values.

\begin{figure}
\includegraphics[width=.49\textwidth]{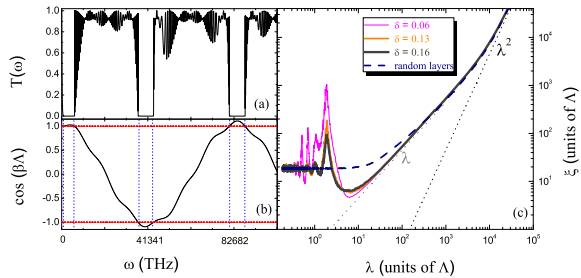}
\caption{(a) The transmission coefficient $T$ and
(b) the parameter $\cos (\beta \Lambda)$ versus the frequency
$\omega$ for the asymmetric periodic M stack described in the main
text (99 layers) and (c) the corresponding localization length
$\xi$ versus the wavelength $\lambda$ for different disorder
parameters $\delta$ (50000 layers).}\label{fig10}
\end{figure}

Finally, let us now consider an asymmetrical M stack where the
condition $|n_1 L_1|=|n_2 L_2|$ is no longer satisfied. In Fig.\
\ref{fig10} we have represented (a) the transmission coefficient
$T$ and (b) the parameter $\cos (\beta \Lambda)$ versus the
frequency $\omega$ for a periodic M stack formed by 50 layers of
length of length $L_1=$ 52.92 nm and index of refraction $n_1=$
-1.58 alternating with 49 layers of length $L_2=$ 28.80 nm and
$n_2=$ 2.12. Note the strong difference between this transmission
spectrum and the symmetrical one (see Fig.\ \ref{fig7}(a)) where a
set of periodically distributed Lorenztian resonances exists.
Despite this fact, the localization length $\xi$ shown in Fig.\
\ref{fig10}(c) presents a region of linear dependence with the
wavelength, as in the symmetric case. However, Eq.\
(\ref{longlocM}) cannot be used to evaluate the localization
length in this region.

\section{Discussion and Conclusions}

We have analyzed numerically the localization length of light
$\xi$ for homogeneous and mixed stacks of layers with index of
refraction $\pm|n_1|$ and thickness $L_1$ alternating with layers
of index of refraction $|n_2|$ and thickness $L_2$. The positions
of the layer boundaries have been randomly shifted with respect to
ordered periodic values. The refraction indices $n_1$ and $n_2$
present no disorder.

For H stacks, the parabolic behavior of the localization length in
the limit of long-wavelengths, previously found in purely
disordered systems \cite{SHENG86,SHENG90,STER93,SHENG95}, has been
recovered. On the other hand, the localization length $\xi$
saturates for very low values of $\lambda$. The transmission bands
modulate the localization length $\xi$ and this modulation
decreases with increasing disorder. Moreover, the localization
length is practically independent of the disorder $\delta$ at the
first gap, that is, it has a very low tendency in this region. We
have also characterized $\xi$ in terms of the reflection
coefficient of alternating layers $|r|^2$ and the optical path
across one grating period $\Lambda_{\rm op}$. Eq.\
(\ref{longlocHinf}) has been proved to be valid for a wide range
of $|r|^2$ values, that is, from transparent to opaque H stacks.
It has also been shown (see Fig.\ \ref{fig5}) that the
localization length $\xi$ is enhanced at the center of each
allowed band.

When left-handed metamaterials are introduced in our system, the
localization length behavior presents some differences with
respect to the traditional stacks, formed exclusively by
right-handed materials. For low-disordered M stacks and
wavelengths of several orders of magnitude greater than the
grating period $\Lambda$, the localization length $\xi$ depends
linearly on $\lambda$ with a slope inversely proportional to the
modulus of the reflection amplitude between alternating layers
$|r|$ (see Eq.\ (\ref{longlocM})). As in the H case, $\xi$
saturates for low-wavelengths, being this saturation constant
proportional to the inverse of $|r|^2$.

If we take into account losses, there is an absorption term whose
absorption length $\xi_{\rm abs}$ is \cite{ASA10(a)}
\begin{equation}\label{abs}
\xi_{\rm abs} = \frac{\lambda}{2\pi\sigma},
\end{equation}
where $\sigma$ is an absorption coefficient. The inverse of the
total decay length is the sum of the inverse of the localization
length $\xi$ plus the inverse of the absorption length $\xi_{\rm
abs}$. Note that $\xi_{\rm abs}$ is proportional to $\lambda$, so,
for low-disordered M stacks and weak absorption metamaterials, the
final expression for the localization length $\xi$ in the linear
region can be written as
\begin{equation}\label{longlocMabs}
\xi = \frac{\lambda}{6 \Lambda_{\rm op} |r| + 2\pi\sigma}.
\end{equation}

In the case of both homogeneous and mixed stacks with
non-symmetric band distribution, that is, when the condition $|n_1
L_1| = |n_2 L_2|$ is not satisfied, the localization length $\xi$
presents an irregular form in all the transmission spectrum. These
changes in $\xi$ are more sensitives in mixed stacks than in
homogeneous structures.

\acknowledgements

The authors would like to acknowledge Vladimir Gasparian for many
interesting discussions. M.O. would like to acknowledge financial
support from the Spanish DGI, project FIS2009-13483.

\bigbreak

\end{document}